\documentclass[prd,aps,showpacs,preprintnumbers,eqsecnum,amsmath,amssymb,nofootinbib]{revtex4}
\usepackage{bm} 
\usepackage{epsfig}

\newcommand{\evr}[2]{\left <\, #1\,\right >_{#2}}
\newcommand{\ever}[1]{\left <\, #1\,\right >}
\newcommand{\matrixel}[3]{\left <\, #1 \left |\, #2\,\right |\, #3\,\right >}

\begin{document}
\title{Wigner's inequalities in quantum field theory} 
\author{
Nikolai Nikitin
} 
\affiliation{
Institute of Nuclear Physics, Moscow State University, 119992, Moscow, Russia}
\author{
Konstantin Toms}
\affiliation{
University of New Mexico, Albuquerque, NM 87131, USA}

\date{\today}
\begin{abstract}


We present a relativistic generalization of the Wigner inequality for the scalar and 
pseudoscalar particles decaying to two particles with spin (fermions and photons.) We 
consider Wigner's inequality with the full spin anticorrelation (with the nonrelativistic 
analog), as well as the case with the full spin correlation. 
The latter case may be obtained by a special choice of the plane of 
measurement of the spin projections on the direction of propagation of fermions.
The possibility for relativistic testing of Bohr's complementarity principle is shown.

\end{abstract}

\maketitle
\section{Introduction}


Since the inception of quantum mechanics in the first quarter of the 20th century, disputes
abound around two closely related issues:\\
\textbf{1)} Is the probabilistic nature of predictions of quantum theory and the 
confirmation by experimental measurements a reflection of the objective laws of the microcosm,
or is the indeterminism a consequence of our ignorance of some ``subtle interactions'' among 
microparticles that would provide theoretical predictions and experimental measurements, such as 
in the case of deterministic classical mechanics? For example, in addition to the well-known measurable 
properties of elementary particles like mass, charge, spin, lepton and baryon numbers, color,
weak isospin, etc., particles may have properties, which, in principle, cannot be measured
with macroscopic analyzers. This lack of information about the values of these variables 
makes the predictions of quantum mechanics probabilistic. This concept is known as
\textit{the theory of hidden variables} of quantum mechanics.\\ 
\textbf{2)} Are the particle parameters described by noncommuting operators elements
of a physical reality \textit{simultaneously and independently of the act of measurement},
or are they fundamentally inseparable from the design and capabilities 
of a particle detector as it is postulated by the Bohr's complementarity principle?

These issues are essential not only for nonrelativistic quantim mechanics (NQM) under which 
they were intensely debated (a comprehensive review may be found in \cite{epr-books}, \cite{found-sing}), but also for 
quantum field theory (QFT). In the framework of QFT this topic was highlighted in a few articles  
(e.g., \cite{sw1987} -- \cite{hiesmayr}). A more complete bibliography may be found in these works.

The experimental answer to the second of the above issues may be given by Bell's inequalities.  
They were introduced for the first time by J.~S.~Bell in 1964--1966 \cite{bell}
and then modified by Clauser, et al. in 1969 \cite{chsh}. 
In Bell's original work, three dichotomic\footnote{Having a spectrum of only two values, in 
this case $\pm 1$.} variables $A$, $B$, and $C$ were introduced. These variables were elements of 
the physical reality simultaneously due to the existence of some set of hidden variables $\lambda$.
The expected values of these dichotomic variables satisfy the following inequality:
\begin{eqnarray}
\label{bell-classic}
\left |\ever{AB}\, -\,\ever{AC} \right |\,\le\, 1\, +\, \evr{BC}. 
\end{eqnarray}
The Clauser-Horne-Shimony-Holt (CHSH) inequality for four
dichotomic variables with spectrum $\pm 1$ is written as follows: 
\begin{equation}
\label{chsh-classic}
\left |\ever{AB}\, +\,\ever{AC}\, +\,\ever{DB}\, -\,\ever{DC} \right |\,\le\, 2. 
\end{equation}
The dichotomic variables $A$, $B$, $C$, and $D$ may be naturally implemented in the form of spin-$1/2$ 
projection in any nonparallel direction $\vec a$, $\vec b$, $\vec c$, or $\vec d$. However, from
the experimentalist's point of view, it is more feasible to use photon polarization and ``flavor--$CP$''
quantum numbers of neutral $K$ and $B$ mesons. For example, there is a recent paper by the
Belle Collaboration on a precise test of Bell's inequalities in neutral $B$ mesons \cite{bell-belle}.

It is widely believed that for the derivation of Bell's inequalities (\ref{bell-classic}) 
and (\ref{chsh-classic}), the existence of a local, context--dependent 
hidden variable $\lambda$ is required. Thus, the violation of Bell's inequalities is often 
considered as a disproof of the existence of a wide class of hidden variables. This view 
comes from classical work \cite{bell}. However, this view is wrong. It was shown in 
\cite{muynck1986} 
that for the derivation of (\ref{bell-classic}) and (\ref{chsh-classic}) it is enough for 
only the non-negative joint probabilities $W(A, B, C)$ and $W(A, B, C, D)$ to exist. 
These probabilities should satisfy Kolmogorov's probability axioms. The existence of such 
probabilities is a mathematical reflection of the following statement: 
\textit{$(A, B, C, D)$ parameters of a given quantum system are simultaneously elements of 
physical reality}. In addition,
we can assume that the existence of the nonnegative joint probabilities is provided
by the hidden variables. Again we emphasize the fact that this assumption
is not necessary for the derivation of formulas (\ref{bell-classic}) and 
(\ref{chsh-classic}).  

From these considerations it follows that Bell's inequalities open the possibility 
for a direct experimental test of the Bohr's complementarity principle. 
In NQM the violation of the inequalities (\ref{bell-classic}) and (\ref{chsh-classic})
may point toward the fact that the observables $A, B, C$, and $D$ (which have the
corresponding noncommuting operators) are simultaneously the elements of the physical
reality. However, this violation may as well point toward the fact that the NQM is a nonlocal
theory (i.e., the measurement of the state of the system in one point will instantly lead
to the change of the state of the system in a different point). J.~Bell adheres to the 
latter opinion \cite{bell-speak}. From the physical point of view the existence of
nonlocality in the NQM follows from the infinite speed of interaction.
                                                                                  
However, in our world the interaction speed is limited by the speed of light. That is why
the degree of the violation of the Bell's inequalities in experiment must be compared 
not with the predictions of NQM (where the inequalities (\ref{bell-classic}) and 
(\ref{chsh-classic}) are violated when we choose some particular directions of the spin
projections), but with the predictions of QFT. In QFT there is a principal difference. 
It is well known \cite{bogolyubov} that the relativistic scattering theory may be built
if we apply to the $S$ matrix the following restrictions: relativistic covariance, unitarity, 
and \textit{causality}. The latter (for example, in the Bogoliubov formulation of the 
principle of casuality in differential form) points toward the independence of $S$ matrix 
behavior in areas separated by spacelike intervals. So the generalization and the study 
of the Bell's inequalities in the framework of the relativistic theory open the possibility 
of the direct experimental test of the Bohr's complementarity principle, free from any 
nonlocality. 

Bell's inequalities in forms (\ref{bell-classic}) or (\ref{chsh-classic}) are not 
suitable enough for relativistic generalization. First, wave functions are used
for the derivation, so it cannot be used in QFT.\footnote{Note that there is a set of
papers in which the authors try to construct the state vector of an entangled system (see, 
for example, \cite{hiesmayr} and references therein). However, such efforts do not lead to 
self-consistent results. It this article we show that this problem may be solved by 
constructing a particular decay Lagrangian, which should be relativistically covariant and
should contain the entanglement by design.} Second, the operators, 
corresponding to $A$, $B$, $C$, $D$ values (usually the particles spin operators), 
should be generalized themselves in a relativistic case. It is desirable to find
a variant of Bell's inequalities without the preceding difficulties. Such variant was 
proposed by E.~Wigner \cite{wigner} in 1970. The Wigner inequalities bind the corresponding
probabilities of the spin states of correlated particles. The procedure of calculation of
such probabilities are well defined in both NQM and QFT. The correlators from the inequalities
(\ref{bell-classic}) or (\ref{chsh-classic}) do not allow such a generalization.
                                                         
Currently there are no suggestions for how to test the principle of complementarity in
the relativistic area for particles with nonzero masses. It is natural to
try to perform such tests on current colliders in the reactions or decays of elementary
particles. For these tests, proper relativistic generalization of Bell's inequalities should be 
introduced for particular processes. Usually elementary particles
are used for such tests as in the decay $\eta_c \to \Lambda \bar\Lambda$ (see \cite{baranov2008}.)
The overview of the main ideas for testing Bell's inequalities in
high-energy physics (HEP) may be found in \cite{bell-hep}. 

In the another set of publications \cite{sw1987} -- \cite{oqs2002}, Bell's inequalities are
studied in the framework of a formal algebraic quantum field theory (AQFT). In this
approach the value of a maximum possible violation of (\ref{chsh-classic}) in QFT 
\cite{sw1988} was found. Also, it was shown that the correlation between the entangled
particles remains even after the local measurement of one of the particles \cite{ch2000}.
This fact, though, is quite obvious because the signal propagation speed in QFT
is limited by the speed of light. However, in AQFT there is no particular suggestion
for testing of these predictions.

In this article we attempt to write a relativistic generalization of Bell's inequalities 
for specific decays of elementary particles. It turns out that the most natural
way for relativistic generalization of Bell's inequalities is Wigner's form \cite{wigner},
which is not dependent on the normalization of states and allows a direct test
of Bohr's complementarity principle in the relativistic region. The spin projections of 
photons and relativistic fermions to various directions were chosen as the observables
with non-commuting operators. Note that it is convenient to use a relativistic
generalization of spin 1/2 from \cite{stech}.  

There is a widely discussed class of time-dependent Bell's inequalities in the phase 
space. Such inequalities may be written using non-negative Wigner functions \cite{new_bell}.
In \cite{leonhardt} it is shown that those inequalities may be used for a case 
of entangled photons for a quantum optical experiment. The contemporary review of this
topic and the corresponding references may be found in \cite{casado}. However, that class
of inequalities is not suitable for testing in HEP experiments. Also, the derivation
of such inequalities is not successive (though correct) in the framework of QFT.

The article is organized as follows: in Sec. \ref{sec:wigner} a few variants 
of Bell's inequalities are obtained. These are suggested for testing the principle of 
complementarity in QFT. In Secs. \ref{sec:ps2ff} and \ref{sec:s2ff}
these inequalities are applied to decays of a scalar and pseudoscalar particle into
a fermion-antifermion pair. Bell's inequalities for the decays into two photons in 
final state are presented in Sec. \ref{sec:gammagamma}.

Some definitions and calculations can be found in the appendixes.


\section{Bell's inequalities in Wigner form}
\label{sec:wigner}

In this section we show that Bell's inequalities in Wigner form may 
be written in two different forms. The first form corresponds to the 
decay of a pseudoscalar particle into a fermion-antifermion 
pair or into two photons. It coincides with \cite{wigner} for
nonrelativistic QM. The second form corresponds to the decay of a 
scalar particle into a fermion and an antifermion or into a photon
pair.  This variant is usually not considered in NQM due to some 
natural obstacles. 

\subsection{Bell's inequalities for two-body decays of pseudoscalar particle}
\label{sec:wigner2}
Let us consider the decay of a resting particle with mass $M$ to a fermion-antifermion pair, 
where we label the momentum of the antifermion as $\vec{k_1}$,  the momentum of the fermion as
$\vec{k_2}$, and their masses as $m_1$ and $m_2$, respectively. Then $\vec{k_1} = -\vec{k_2}$
and $M > m_1 + m_2$. If the decay is induced by the strong or electromagnetic
interaction, the flavors of fermions are conserved ($m_1 = m_2 = m$), 
as well as $P$ parity ($P_{f\bar{f}} = (-1)^{L_{f\bar{f}}+1} = -1$). The full momentum
of the system is conserved, so $J_{f\bar{f}} = 0$. Then the orbital momentum and the
spin of the fermion-antifermion pair is $L_{f\bar{f}} = S_{f\bar{f}} = 0$. This leads to the
full anticorrelation of the fermion ``2'' spin and the antifermion ``1'' spin projections 
in any direction determined by the vector $\vec{a}$:
\begin{eqnarray}
\label{s-anticorr}
s^{(2)}_a = -s^{(1)}_a
\end{eqnarray}
We stress the fact that in local QFT the perfect anticorrelations for spin 
do appear only in the moment of the decay and only in the point of the decay.
These anticorrelations may be described using the corresponding matrix elements, as
we show in the Secs. \ref{sec:ps2ff} and \ref{sec:gammagamma}. If all of the 
interactions of the anticorrelated particles with any of the external particles or 
fields may be described in terms
of the perturbation theory, then the violation of the perfect anticorrelation may
be considered as small, and we may consider (\ref{s-anticorr}) to hold true everywhere. 
Let us prove that statement. Actually, the charged fermions or photons with a typical
energy of a few GeV interact with the environment mainly by the laws of QED,
where the coupling constant $\alpha_{em} \approx 1/137$. So one can expect that in real 
experiments the violation of (\ref{s-anticorr}) due to nonlocality will be at the level of 
a few percent (and this fact is also well-known from the various calculations in 
the QED framework). As we show in what follows, the level of the violation of all the considered
Wigner inequalities is much higher than the aforementioned effects, and these few-percent 
contributions to the decay amplitudes are not able to cancel the violation of the 
inequalities. 

Next, suppose that the fermion and antifermion spin projections on three nonparallel
directions $\vec{a}$, $\vec{b}$, and $\vec{c}$ are the elements of the physical reality
at the same time. In Appendix A.1 it is shown that such an assumption leads to the
inequality 
\begin{eqnarray}
\label{wigner-ps}
w\left (s^{(2)}_a = +\,\frac{1}{2},\, s^{(1)}_b = + \,\frac{1}{2} \right )\,& \le & 
w\left (s^{(2)}_a = +\,\frac{1}{2},\, s^{(1)}_c = + \,\frac{1}{2}\right )\, +  
w\left (s^{(2)}_c = +\,\frac{1}{2},\, s^{(1)}_b = + \,\frac{1}{2}\right )
\end{eqnarray}
for the probabilities for fermion and antifermion to have spin projection of +1/2 (at the
same time) on any two of three directions $\vec{a}$, $\vec{b}$, and $\vec{c}$. 
Since only the decay probabilities appear in (\ref{wigner-ps}), it is equally applicable in 
QFT and NQM. Let all the vectors $\vec{a}$, $\vec{b}$, and $\vec{c}$ lie on the same 
$XZ$ plane. In nonrelativistic QM this leads to the transformation of (\ref{wigner-ps}) 
into the following inequality (see \cite{wigner}):
\begin{eqnarray}
\label{wigner-classic-nqm}
\sin^2\frac{\theta_{ab}}{2}\,\le\,\sin^2\frac{\theta_{ac}}{2}\, +
                                \,\sin^2\frac{\theta_{bc}}{2} 
\end{eqnarray}
where $\theta_{\alpha \beta} = \theta_{\alpha} - \theta_{\beta}$ and $\{\alpha, \beta \} = 
\{a, b, c\}$. The inequality (\ref{wigner-classic-nqm}) is violated when vectors 
$\vec{a}$ and $\vec{b}$ form an angle less than $\pi$  and vector $\vec{c}$ 
bisects this angle. The maximal violation of (\ref{wigner-classic-nqm}) is achieved for
$\theta_{ab}=2\pi/3$ and $\theta_{ac} = -\theta_{bc} = \pi/3$. 
The evidence of the violation of (\ref{wigner-classic-nqm})
is a direct experimental confirmation of the Bohr's complementarity principle.
It is possible to write an inequality analogous to (\ref{wigner-ps}) for the neutral
resting particle decays into two pions, for example, $\pi^0 \to 2 \gamma$. In this case
the system of two photons with negative $P$ parity has the full momentum of zero, as
well as orbital momentum and spin; that is, 
$
J_{\gamma\gamma} = L_{\gamma\gamma} = S_{\gamma\gamma} = 0
$. This implies that for a photon with linear polarization, there is a full anticorrelation of
polarization in any direction $\vec{a}$, which is perpendicular to the direction of
the photon propagation. If we label the polarization of photon with $\lambda^{(1,2)}_{\alpha}$
and the corresponding states with indices ``1'' and ``2'',  then the analog of (\ref{wigner-ps})
for photons will look like
\begin{eqnarray}
\label{wigner-foton-ps}
    w\left (\lambda^{(1)}_a = 1,\, \lambda^{(2)}_b = 1 \right )\, \le  
    w\left (\lambda^{(1)}_a = 1,\, \lambda^{(2)}_c = 1 \right )\, + \,  
    w\left (\lambda^{(1)}_c = 1,\, \lambda^{(2)}_b = 1\right ).
\end{eqnarray}

We note that the inequalities (\ref{wigner-ps}) and (\ref{wigner-foton-ps}) (as well as the 
inequality (\ref{bell-classic})) require an ideal experimental situation with the full 
anticorrelation (\ref{s-anticorr}). In a real experiment, some corrections must be introduced
in order to take into account the efficiencies of the detectors. Also, we note that the 
inequality (\ref{chsh-classic}) in principle does not have such difficulties, because it
takes into account contributions from both correlations and anticorrelations. The
mere derivation of this fact can be found, for example, in \cite{found-sing}.

 
\subsection{The condition of the full spin correlation in the decays of a pseudoscalar particles
and a new expression for Bell's inequalities}
\label{sec:wigner_angles}
Let us consider the decay of a resting scalar particle ($S$) with mass $M$ into a fermion-antifermion
pair. In the case of strong or electromagnetic decay, $P$ parity is conserved, leading to the
condition $L_{f\bar f} = S_{f\bar f} = 1$ for $J_{f\bar f}=0$.
It is easy to see that in this case, if the projection of the full spin of the $f\bar f$ pair on the
direction $\vec{a}$ is equal $S^{f\bar f}_a = \pm 1$, then full correlation exists between the spin projections of the 
fermions on this direction:
\begin{eqnarray}
\label{s-corr}
s^{(2)}_a\, =\, s^{(1)}_a.
\end{eqnarray}
The arguments in favor of using (\ref{s-corr}) are exactly the same as
the preceding arguments for using (\ref{s-anticorr}).
In the case of $S^{f\bar f}_a = 0$, the full anticorrelation exists instead, like in the case of the decay of 
a pseudoscalar particle. With arbitrary orientation of the plane of measurement, both cases may
take place simultaneously -- the correlation and the anticorrelation, and the experimental possibility
of the testing of the Bell's inequalities will suffer.
However, it is possible to choose such a relative position of the measurement plane and the propagation
direction of fermions that the contribution from the anticorrelation
will be insignificant. In such an experimental
configuration it becomes possible to test the Bell's inequalities with full correlation.

Let $\theta$ and $\phi$ be the zenith and azimuth angles of the vector $\vec{n}$, respectively, and
$\tilde\theta$ be the angle between the $\vec{a}$ and $\vec{n}$ vectors. The angle $\theta_a$ defines
the position of the vector $\vec{a}$ in the $XZ$ plane. The
$\tilde\theta$ dependence of the amplitude of the decay of
pseudoscalar meson $S$ into a fermion-antifermion pair is
\begin{eqnarray}
A(S \to f \bar f)\, &\sim&
\matrixel{S_{f \bar f} = 1,  S_a^{f \bar f} = 0}{H}{S_S = 0, S_a^{(S)} = 0}\,
\cos\tilde\theta\, + \nonumber \\ 
&+& \left (
\matrixel{S_{f \bar f} = 1,  S_a^{f \bar f} = + 1}{H}{S_S = 0, S_a^{(S)} = 0}\, -
      \right . \nonumber \\ 
&-& \left .
e^{2 i \phi}\matrixel{S_{f \bar f} = 1,  S_a^{f \bar f} = - 1}{H}{S_S = 0, S_a^{(S)} = 0}
      \right )\sin\tilde\theta. \nonumber 
\end{eqnarray} 
One can see that the anticorrelations are gone in the case of angle $\tilde\theta = \pi/2$.
Then, using the cosine theorem: 
\begin{eqnarray}
0\, =\,\cos\theta\,\cos\theta_a\, +\,\sin\theta\,\sin\theta_a\,\cos\phi.\nonumber
\end{eqnarray}
When testing the Bell's inequalities in Wigner's form for the full spin correlation, it is necessary to choose 
such values of the angles $\theta$ and $\phi$ that the cosine theorem hold for the angle $\theta_a \in [0, \pi]$.
This is possible when
\begin{eqnarray}
\label{pi-popolam}
\theta\, =\,\phi\, =\,\tilde\theta\, =\,\pi/2, 
\end{eqnarray}
that is, in the case when the fermions propagate in the direction of the axis $Y$, perpendicular 
to the polarization measurement plane $XZ$. When the position of the polarizers in the $XZ$ plane
is fixed, there is only one direction (\ref{pi-popolam}) for which the condition of the full 
correlation (\ref{s-corr}) is fulfilled for every direction $\vec{a}$ in the $XZ$ plane. This
configuration is very special and requires an additional experimental selection procedure.

Later we always assume such an experimental 
configuration when talking about the decays of the scalar particles.
Again, as in the Sec. \ref{sec:wigner2}, suppose that spin projections of fermion and 
antifermion on any of three nonparallel directions $\vec a$, $\vec b$, and $\vec c$ are 
simultaneously elements of the physical reality. Then we can obtain the following equation 
for the full correlation (is true for the condition (\ref{pi-popolam})):
\begin{eqnarray}
\label{wigner-s}
w\left (s^{(2)}_a = +\,\frac{1}{2},\, s^{(1)}_b = - \,\frac{1}{2} \right )\,& \le & 
w\left (s^{(2)}_a = +\,\frac{1}{2},\, s^{(1)}_c = - \,\frac{1}{2}\right )\, + \, 
w\left (s^{(2)}_c = +\,\frac{1}{2},\, s^{(1)}_b = - \,\frac{1}{2}\right ) 
\end{eqnarray}
derived in Appendix \ref{appsub:wigner-corr}. 
The inequality (\ref{wigner-s}) is not considered in NQM.

In the case of the decay of a scalar particle into two photons, for example $H^0 \to \gamma \gamma$, 
in the particle's rest frame, there appears a two-photon state with 
$ 
J_{\gamma\gamma} = 0 
$ 
and 
$ 
L_{\gamma\gamma} = S_{\gamma\gamma} = 2 
$. 
For this state, the full spin correlation is possible when 
$S^{\gamma \gamma}_a = \pm 2$, as well as the full anticorrelation when 
$S^{\gamma \gamma}_a = 0$, even with the condition (\ref{pi-popolam}). The probability of 
correlation is proportional to 
$ 
2\,\left | Y^2_2 (\theta = \pi/2,\, \phi = \pi/2)\right |^2 = 15/ 16 \pi 
$, 
while the probability of anticorrelation is: 
$ 
\left | Y^0_2 (\theta = \pi/2,\, \phi = \pi/2)\right |^2 = 5/16 \pi 
$, 
where $Y^m_\ell (\theta,\, \phi)$ are spherical functions. 

If the photon polarizations are fully correlated, then for such case it is possible to write 
the inequality analogous to the inequality (\ref{wigner-s}) for the case of the full anticorrelation: 
\begin{eqnarray} 
\label{wigner-foton-s} 
    w\left (\lambda^{(1)}_a = 1,\, \lambda^{(2)}_b = 2 \right )\, \le  
    w\left (\lambda^{(1)}_a = 1,\, \lambda^{(2)}_c = 2 \right )\, + \,  
    w\left (\lambda^{(1)}_c = 1,\, \lambda^{(2)}_b = 2\right ). 
\end{eqnarray} 
Considering the probabilities for the correlation and the anticorrelation above 
$\theta = \phi = \pi/2$, it is possible to write the Bell's inequality for the decay 
$H^0 \to \gamma \gamma$: 
\begin{eqnarray}
\label{wigner-foton-ps-s} 
  3\, w\left (\lambda^{(1)}_a = 1,\, \lambda^{(2)}_b = 2 \right )\, 
&+&   w\left (\lambda^{(1)}_a = 1,\, \lambda^{(2)}_b = 1 \right )\,\le \\ 
&\le &  
  3\, w\left (\lambda^{(1)}_a = 1,\, \lambda^{(2)}_c = 2 \right )\, + \, 
      w\left (\lambda^{(1)}_a = 1,\, \lambda^{(2)}_c = 1 \right )\, +  \nonumber\\  
&+& 
  3\, w\left (\lambda^{(1)}_c = 1,\, \lambda^{(2)}_b = 2 \right )\, +\, 
      w\left (\lambda^{(1)}_c = 1,\, \lambda^{(2)}_b = 1\right ). \nonumber 
\end{eqnarray} 

We again stress the fact that the inequalities (\ref{wigner-s}) and (\ref{wigner-foton-ps-s})
require an even ``more ideal'' experimental procedure because we require not only the corellation
(\ref{s-corr}) but also the special configuration (\ref{pi-popolam}).  


 

\section{Bell's inequalities in QFT for the decay of a pseudoscalar particle into two fermions}
\label{sec:ps2ff}
In the case of conserved $P$ parity the decay of a pseudoscalar particle to a fermion-antifermion 
pair can be described using an effective Hamiltonian:
\begin{eqnarray}
\label{Heff_for_PS2ff}
\mathcal{H}^{(PS)}(x)\, =\, g\,\varphi (x)\,\left (\bar f(x)\,\gamma^5\, f(x)\right )_N, 
\end{eqnarray}
where $g$ is the effective coupling constant, $\varphi (x)$ is the field of the pseudoscalar 
particle, $f(x)$ is the fermionic field, and the $\gamma^{5}$ matrix is defined in the Appendix 
\ref{app:spin}. The fermionic current is written in the normal form
(reflected by the $N$ index).
In the decays described by (\ref{Heff_for_PS2ff}), 
the masses of fermion and antifermion should be equal. In all
subsequent equations we use the index ``1'' for antifermion and the
index ``2'' for fermion (similarly for their masses).
The unitary normal vector (\ref{n-1v}) coincides with the fermion
propagation direction, that is, 
$
\vec k_2\, =\, |\vec k_2|\,\vec n
$.
Let us sequentially consider three possible cases.


\subsection{The decay of a resting pseudoscalar particle}
\label{sub:ps1}

Let the pseudoscalar be at rest at the origin of the coordinate system 
and place a spin state analyzers at infinity to measure spin projections in planes
parallel to the $XZ$ plane. If the spin projections of the fermion on the $\vec a$ direction and 
of the antifermion on the $\vec b$ direction are equal to $+ 1/2$, 
the decay amplitude can be written as follows:
\begin{eqnarray}
A\left (s^{(2)}_a = +\,\frac{1}{2},\, s^{(1)}_b = + \,\frac{1}{2} \right ) &=& 
-\, g\,\bar u(\vec k_2, s^{(2)}_a = + 1/2, \vec a\, )\,\gamma^5\, 
     v(\vec k_1, s^{(1)}_b = + 1/2, \vec b\, )\, = \nonumber \\
&=& g\,\sqrt{\frac{\varepsilon_2 + m_2}{\varepsilon_1 + m_1}}\, 
\left (M + m_1 - m_2 \right )\,\chi^{\dagger}_+(\vec a\, )\,\chi_-(\vec b\, ), 
\nonumber 
\end{eqnarray}
where the four-component spinors $u$ and $v$ are defined as
(\ref{dirac_u}) and (\ref{dirac_v}), respectively.
This amplitude does not depend on angular variables. Then, taking into account the first equation 
of (\ref{formuli-chi}), the probability can be written in the following way:
\begin{eqnarray}
\label{w_ab++}
w\left (s^{(2)}_a = +\,\frac{1}{2},\, s^{(1)}_b = + \,\frac{1}{2} \right )\, =\,
f(M, \, m_1,\, m_2,\, \theta,\, \phi)\, \sin^2\frac{\theta_{ab}}{2}.
\end{eqnarray}
If the direction of fermion propagation is not taken into account, then
\begin{eqnarray}
\label{f1} 
f(M, \, m_1,\, m_2,\, \theta,\, \phi)\, =\,\frac{g^2}{16\,\pi}\, 
\frac{M^2 - (m_1 - m_2)^2}{M^3}\,\lambda^{1/2}(M^2,\, m^2_1,\, m^2_2),  
\end{eqnarray}
where $\lambda (a, b, c) = a^2 + b^2 + c^2 - 2ab -2ac -2bc$ is the triangular function, 
defining the dependency of the probability on the phase space.

From (\ref{w_ab++}) and (\ref{f1}) it follows that in the framework of QFT the
Bell's inequality (\ref{wigner-ps}) reduces to (\ref{wigner-classic-nqm}), which was obtained
in the nonrelativistic approach. This result follows from the factorization of the spin
part of the amplitude 
$A\left (s^{(2)}_a = +\,1/2,\, s^{(1)}_b = + \,1/2 \right )$ 
due to zero orbital momentum in this decay.  

If we take into account the fermion propagation direction, that is, select only the fermions with
fixed values of $\tilde\theta$ and $\tilde\phi$, then Eq. (\ref{f1}) should be modified as
\begin{eqnarray}
\label{f2} 
f(M, \, m_1,\, m_2,\, \tilde\theta,\, \tilde\phi)\, =\,\frac{g^2}{16}\, 
\frac{M^2 - (m_1 - m_2)^2}{M^3}\,\lambda^{1/2}(M^2,\, m^2_1,\, m^2_2)\, 
\frac{\sin\tilde\theta}{(2\,\pi)^2}.  
\end{eqnarray}
Equation (\ref{f2}) reflects the fact that if both fermion and antifermion propagate along the
$z$ axis (i.e. $\sin\tilde\theta = 0$), they would not be registered and inequality (\ref{wigner-ps}) 
becomes meaningless. We emphasize the fact that in the nonrelativistic approach, this obvious deduction
cannot be made.




\subsection{The adjustment due to the non-antiparallelity of the fermion and antifermion momenta}
\label{sub:ps3}

In the previous subsection we showed that, due to the nonconservation of the momentum projection on 
the $y$ axis in the rest frame of a pseudoscalar particle, the angle between the vectors $\vec k_2$ 
and $\vec k_1$ had a small deviation from $\pi$. Due to the violation of antiparallelity of the two vectors in
the Bell's inequality (\ref{wigner-classic-nqm}), a quadratic correction appeared by the small 
parameter $k_y/M$. 

The antiparallelity of the vectors $\vec k_2$ and $\vec k_1$ can be caused by
the emission of a soft photon from one of the fermions. The energy $\omega$ of the
soft photon can be below the detection threshold and a lot less than energies $\varepsilon_1$ and $\varepsilon_2$ 
(these are of the order of $M/2$). It is well known from standard QED that 
in the first approximation by $\omega /\varepsilon_{1,2}$ the amplitude of the emission
of a soft photon can be factorized by the amplitude for the no-emission process and by a factor 
corresponding to the emission of a soft photon. Therefore, in the case of soft photon emission
the possible corrections for (\ref{wigner-classic-nqm}) may only appear starting in the second
order by the small parameter 
$
\omega /\varepsilon_{1,2} \sim \omega /M
$.

Let us prove the general statement. Let
$
\vec k_1 = |\vec k_1| \vec n_1
$ 
and 
$
\vec k_2 = |\vec k_2| \vec n
$ 
and the conservation law
\begin{eqnarray}
\label{n1-n-ell}
|\vec k_1|\,\vec n_1\, +\,|\vec k_2|\,\vec n\, =\, |\vec p\, |\,\vec \ell, 
\end{eqnarray} 
where the vector $\vec \ell$ in not parallel to the vectors $\vec n_1$ and $\vec n$. Additionally, let
$E = \varepsilon_1 + \varepsilon_2$ and
\begin{eqnarray}
\frac{|\vec p\, |}{M}\,\ll\, 1. \nonumber
\end{eqnarray}
Then
\begin{eqnarray}
\label{main-O}
w\left (s^{(2)}_a = +\,\frac{1}{2},\, s^{(1)}_b = + \,\frac{1}{2} \right )\, =\, 
g^2\, f_0 (M, m_1, m_2)\,\sin^2 \frac{\theta_{ab}}{2}\, +\,
O\left (\frac{|\vec p\, |^2}{M^2} \right). 
\end{eqnarray}

Actually, with zero angular momentum there is a spherical symmetry. Small deviations from
such symmetry may only be quadratic by any direction, which immediately leads to the formula
(\ref{main-O}). However, we also prove this fact strictly.

When the projections of the fermion spin on the direction of $\vec a$ and the antifermion spin on the 
direction of $\vec b$ both are equal to $+ 1/2$, the decay amplitude can be written as follows:
\begin{eqnarray}
\label{A-L}
&&\tilde A\left (s^{(2)}_a = +\,\frac{1}{2},\, s^{(1)}_b = + \,\frac{1}{2} \right )\, =\, 
g\,\Bigl [ \sqrt{\varepsilon_1 + m_1}\,\sqrt{\varepsilon_2 + m_2}\,
\chi^{\dagger}_+(\vec a\, )\,\chi_-(\vec b\, )\, - \nonumber \\
&&\qquad -\,\sqrt{\varepsilon_1 - m_1}\,\sqrt{\varepsilon_2 - m_2}\,\, 
\chi^{\dagger}_+(\vec a\, )\,
\frac{\bigl (\vec \sigma\,\vec k_2\bigr )\,\bigl (\vec\sigma\,\vec k_1\bigr )}
     {\bigl |\vec k_2 \bigr |\,\,\bigl |\vec k_1 \bigr |}
\chi_-(\vec b\, )
\Bigr ].  
\end{eqnarray}

From (\ref{A-L}) and (\ref{n1-n-ell}) it follows that if vectors $\vec n_1$ and $\vec n$ are not 
antiparallel, then the amplitude can be written down in the form\footnote{Later we assume 
$\epsilon^{123} = \epsilon_{123} = +1$.
}
\begin{eqnarray}
\label{A-O}
&&\qquad\quad 
  A\left (s^{(2)}_a = +\,\frac{1}{2},\, s^{(1)}_b = + \,\frac{1}{2} \right ) \, =\,
g\,\sqrt{\frac{\varepsilon_2 + m_2}{\varepsilon_1 + m_1}}\\ 
&& \left [
\left ( E\, +\, m_1\, -\, m_2\, -\, 
\sqrt{\frac{\varepsilon_2 - m_2}{\varepsilon_2 + m_2}}\,\, |\vec p\, |\,\ell^i n^i 
\right )\,\sin \frac{\theta_{ab}}{2}\, -
\, i\,\sqrt{\frac{\varepsilon_2 - m_2}{\varepsilon_2 + m_2}}\,\, 
|\vec p\, |\,\epsilon^{ijk} n^i \ell^j w^k_{++} \right ].\nonumber  
\end{eqnarray}

The series expansion by the small parameter $|\vec p\, |/M$ gives
\begin{eqnarray}
\label{O2}
&&E\, =\, E^{(0)}\, +\, E^{(1)}\,\frac{|\vec p\, |}{M}\,\ell^i n^i\, +\,
       O\left (\frac{|\vec p\, |^2}{M^2} \right);\nonumber \\
&&\varepsilon_{1,2}\, =\,\varepsilon^{(0)}_{1,2}\, +\, 
                     \varepsilon^{(1)}_{1,2}\,\frac{|\vec p\, |}{M}\,\ell^i n^i\, +\, 
                     O\left (\frac{|\vec p\, |^2}{M^2} \right); \\
&&d\Phi_n\, =\,\frac{d\Omega}{4\,\pi}
\left (\Phi_2^{(0)}\, +\,  
       \Phi_2^{(1)}\,\frac{|\vec p\, |}{M}\,\ell^i n^i\, +\,
       O\left (\frac{|\vec p\, |^2}{M^2} \right) \right )\, d\Phi_{n-2}, \nonumber
\end{eqnarray}
where the phase space $d\Phi_n$ includes the integration over the variables different from
the angle variables of the fermion and considers a possible emission of $(n-2)$ soft
photons, while $d\Omega = d\cos\theta d\phi$ selects the integration over the fermion propagation direction.
The explicit form of the coefficients of the expansion in (\ref{O2}) depends on the source of the momentum $\vec p$.
For example, if that momentum results from the Brownian motion 
of the decaying pseudoscalar particle, then
$
E\,\approx\, M\,\left (1\, +\,\frac{1}{2}\,\frac{|\vec p\, |^2}{M^2} \right )
$.
Hence $E^{(0)} = M$, $E^{(1)} = 0$.

In accordance with (\ref{A-O}) and (\ref{O2}) for the decay probability, it can be written
\begin{eqnarray}
\label{w-O}
&&w\left (s^{(2)}_a = +\,\frac{1}{2},\, s^{(1)}_b = + \,\frac{1}{2} \right ) \, =\, 
g^2\,\int\, d\Omega_{n-1}\, \int\,\frac{d\Omega}{4\,\pi}\, 
\left (
\alpha^{(0)}\,\sin^2 \frac{\theta_{ab}}{2}\, + \right .\nonumber \\
&& \qquad\qquad\left .  +\, 
\alpha^{(1)}_1\,\frac{|\vec p\, |}{M}\,\ell^i n^i\, +\,
\alpha^{(1)}_2\,\frac{|\vec p\, |}{M}\,\epsilon^{ijk} n^i \ell^j Im \bigl (w^k_{++}\bigr )
\, +\, O\left (\frac{|\vec p\, |^2}{M^2}\right ) 
\right ). 
\end{eqnarray}
Given that
\begin{eqnarray}
\int\,\frac{d\Omega}{4\,\pi}\, =\, 1\qquad\textrm{и}\qquad
\int\,\frac{d\Omega}{4\,\pi}\, n^i\, =\,\evr{n^i}{\Omega}\, =\, 0, \nonumber
\end{eqnarray}
then (\ref{main-O}) immediately follows from (\ref{w-O}).

Note that the statement (\ref{main-O}) is quite general. It is true for the Brownian motion 
of a decaying particle, the uncertainties from the composition of the initial state, the
interaction between the final state fermions (e.g. via the Coulomb force), or the interaction with
a weak external field. In the beginning of the current subsection two cases complying with (\ref{main-O}) 
were presented: the emission of soft photons and a phase-space limit of the decay. The equality (\ref{main-O}) 
may be useful not only for the Bell's inequalities. It can be easily adapted to the various tasks in quantum 
teleportation and quantum measurements.

As an example, let us consider the influence of two parallel spin
analyzers situated in the $XZ$ plane
and crossing the $Y$ axis at the distance $\pm L/2$ from the coordinates' origin. Then the uncertainty
of $k_y \sim 1/L$ and 
$$
\frac{k^2_y}{M^2}\,\sim\,\frac{1}{(M\, L)^2}\,\ll\, 1.
$$ 
For example, if $L \sim $ 2 cm and $M \sim$ 1 GeV, then $1/(M\, L)^2 \sim 10^{-28}$, 
which is below the current available experimental precision by many orders of magnitude.
Thus, if the distance between the spin analyzers is macroscopic, then this adjustment 
is unimportant. Later in this article we always suppose the analyzers reside at infinity.


\section{The Bell's inequalities in QFT for the decay of a scalar particle into two fermions}
\label{sec:s2ff}

The effective Hamiltonian of the decay of a scalar particle to a fermion-antifermion pair
can be written in exactly the same way as (\ref{Heff_for_PS2ff}):
\begin{eqnarray}
\label{Heff_for_S2ff}
\mathcal{H}^{(S)}(x)\, =\, g\,\varphi (x)\,\left (\bar f(x)\, f(x)\right )_N, 
\end{eqnarray}
where $\varphi (x)$ is a field of the scalar particle and the rest of the definitions are the same as for
Eq. (\ref{Heff_for_PS2ff}). Like in Sec. \ref{sec:ps2ff}, the index
``1'' is always for antifermion, and index  ``2'' for fermion. The unitary vector $\vec n$ 
is set to the direction of the fermion propagation, while the vector $\vec n_1$ is set to the antifermion direction. 

We only consider the case when the scalar particle is resting and again spin analyzers are placed
at infinity in the planes parallel to the $XZ$ plane. 


\subsection{Decay of the scalar particle in the case when the direction of fermion propagation is fixed}

As was shown in the Sec. \ref{sec:wigner_angles}, the spin correlations are only possible when $\theta = \phi = \pi/2$.
That is why we consider the decay of a scalar particle into fermion-antifermion pair for fixed
direction.

Let us consider a fermion propagating along some chosen direction defined by angles $\theta$ and $\phi$. 
In this case the angular part of the probability 
$
w\left (s^{(2)}_a = +\,\frac{1}{2},\, s^{(1)}_b = - \,\frac{1}{2} \right )
$
has the form
\begin{eqnarray}
\label{w_theta-phi}
&& \sin\theta\,\left |
\chi^{\dagger}_+(\vec a\, )\,
\bigl ( \vec \sigma\,\vec n \bigr )
\chi_+(\vec b\, ) 
\right |^2\, =\, 
\sin\theta\,
\left (
\sin^2\theta\,\cos^2\phi\,\sin^2\frac{\kappa_{ab}}{2}\, +\right . \nonumber \\ 
&&\qquad\left . \cos^2\theta\,\cos^2\frac{\kappa_{ab}}{2}\, +\,
\frac{1}{2}\,\sin (2\theta)\,\cos\phi\,\sin\kappa_{ab}\, +\, 
\sin^2\theta\,\sin^2\phi\,\sin^2\frac{\theta_{ab}}{2}
\right ). 
\end{eqnarray}
According to the Sec. \ref{sec:wigner_angles}, $\theta = \phi = \pi/2$ in (\ref{w_theta-phi}).
Then,
\begin{eqnarray}
\label{w_phi}
w\left (s^{(2)}_a = +\,\frac{1}{2},\, s^{(1)}_b = - \,\frac{1}{2} \right ) \,\sim\, 
\sin^2\frac{\theta_{ab}}{2},
\end{eqnarray}
and we get the Bell's inequalities in the form (\ref{wigner-classic-nqm}), as for the case of a pseudoscalar 
particle decay. However, in this case the direction is fixed, and much higher experimental statistics is 
required for testing of the inequalities. 
For the decay of the scalar particle to a fermion-antifermion pair, 
small deviations from the antiparallel state of the vectors $\vec n_1$ and $\vec n$ are 
proportional to the first power of $|\vec p\, |\, (\vec\ell\, \vec n\, )$. The result (\ref{w_phi})
is somehow trivial. Actually, in the chosen kinematical configuration the fermions propagate along
the $Y$ axis perpendicular to the $XZ$ plane. The Lorentz transformation along this axis does not 
have any impact on the spin correlations in the $XZ$ plane compared to the nonrelativistic case.


\section{Bell's inequalities for the decays of scalar and pseudoscalar particles into two photons}
\label{sec:gammagamma}

Consider a decay of a particles without spin into two photons in the final state.
In what follows we use a Higgs boson as a scalar and a $\pi^0$ meson as a pseudoscalar.

As long as $P$ parity is conserved, the amplitude of $H^0 \to \gamma\gamma$ decay has the following form:
\begin{eqnarray}
A^{H^0 \to \gamma\gamma}\, =\, F_H\, 
\epsilon^{*\,\mu} (\lambda^{(1)})\,\epsilon^*_{\mu} (\lambda^{(2)}), \nonumber
\end{eqnarray}
where $F_H$ is the known constant which does not affect the result, and the 
$\vec k_2 = \omega \vec n = \omega \left ( 0, 1, 0 \right ) $. The 4-vectors of the photon polarization in the direction $\vec a$, 
orthogonal to $\vec n$, are defined as follows:
\begin{eqnarray}
\epsilon^{\mu} (\lambda^{(1, 2)}_a = 1)\, =\, 
\bigl (
0,\,\sin\theta_a,\, 0,\,\cos\theta_a
\bigr );\qquad
\epsilon^{\mu} (\lambda^{(1, 2)}_a = 2)\, =\, 
\bigl (
0,\,\cos\theta_a,\, 0,\, -\,\sin\theta_a
\bigr ). \nonumber
\end{eqnarray}
In the case when the first photon has polarization ``1'' in the direction $\vec a$ and the 
second photon has polarization ``2'' in the direction orthogonal to $\vec b$, the amplitude can be 
written as follows: 
\begin{eqnarray}
A^{H^0 \to \gamma\gamma}\left (\lambda^{(1)}_a = 1,\, \lambda^{(2)}_b = 2 \right )\, =\, 
F_H\,\epsilon^{*\,\mu} (\lambda^{(1)}_a = 1)\,
     \epsilon^*_{\mu}  (\lambda^{(2)}_b = 2)\, =\, 
-\, F_H\,\sin \theta_{ab}. \nonumber
\end{eqnarray}
In the case when the first photon has polarization ``1'' in the direction $\vec a$ and the 
second photon has polarization ``1'' in the direction $\vec b$, the amplitude can be 
written as follows: 
\begin{eqnarray}
A^{H^0 \to \gamma\gamma}\left (\lambda^{(1)}_a = 1,\, \lambda^{(2)}_b = 1 \right )\, =\, 
F_H\,\epsilon^{*\,\mu} (\lambda^{(1)}_a = 1)\,
     \epsilon^*_{\mu}  (\lambda^{(2)}_b = 1)\, =\, F_H\,\cos \theta_{ab}. \nonumber
\end{eqnarray}
Hence, Bell's inequality (\ref{wigner-foton-ps-s}) reduces to the following trigonometric 
inequality:
\begin{eqnarray}
\label{wigner-nonclassic-foton}
\sin^2\theta_{ab}\,\le\,\frac{1}{2}\, +\,\sin^2\theta_{ac}\, +\,\sin^2\theta_{bc}. 
\end{eqnarray}
It is a new version of Bell's inequality, which in principle may be violated. However, the 
range of the angles where it is violated is very narrow, and it is written with the
condition (\ref{pi-popolam}).

At the Large Hadron Collider, where the Higgs bosons may be born, the following decay is 
possible: $H \to \gamma^* \gamma^* \to (\ell^+ \ell^-)\, (\ell^+ \ell^-)$. In this case, 
each of ($\ell^+\ell^-$) planes may be used as a spin projector for each photon. 
The invariant mass of a lepton pair should be small in order to effectively exclude
the contribution from a highly virtual photon (such photons have additional longitudinal 
polarization). Effects of the exchange interaction are negligible because both lepton 
pairs are strongly separated in the phase space in the rest frame of $H^0$. 

Let us now consider a decay $\pi^0 \to \gamma \gamma$. The amplitude of that decay has the form: 
\begin{eqnarray}
A^{\pi^0 \to \gamma\gamma}\, =\, F_{\pi}\,
\varepsilon_{\mu\nu\alpha\beta} 
\epsilon^{*\,\mu} (\lambda^{(1)})\,\epsilon^{*\,\nu} (\lambda^{(2)})\, 
k^{\alpha}_1\, k^{\beta}_2. \nonumber
\end{eqnarray}
Given that in this problem
$
k^{\mu}_1 = \omega (1, 0, -1, 0)
$, 
$
k^{\mu}_2 = \omega (1, 0, 1, 0)
$,
and 
$
\varepsilon_{0123}\, =\, -\, \varepsilon^{0123}\, =\, + 1
$, 
we have:
\begin{eqnarray}
A^{\pi^0 \to \gamma\gamma}\left (\lambda^{(1)}_a = 1,\, \lambda^{(2)}_b = 1 \right )\, =\,
-\, 2\,\omega^2\, F_{\pi}\,\sin \theta_{ab}. \nonumber
\end{eqnarray}
Substitution of the amplitude into inequality (\ref{wigner-foton-ps}) again leads to
the inequality
\begin{eqnarray}
\label{wigner-classic-foton}
\sin^2\theta_{ab}\,\le\,\sin^2\theta_{ac}\, +\,\sin^2\theta_{bc}, 
\end{eqnarray}
which is violated when the vectors $\vec a$ and $\vec b$ form an \textit{acute} angle, 
while $\vec c$ is its bisector. The maximal violation of the inequality (\ref{wigner-classic-foton}) 
is achieved when $\theta_{ab} = \pi/3$ and $\theta_{ac} = -\theta_{bc} = \pi/6$.
The inequality (\ref{wigner-classic-foton}) 
is analogous to the trigonometric inequality (\ref{wigner-classic-nqm}). 
 
Thus, Bell's inequalities in Wigner form for the decay of the scalar (\ref{wigner-foton-ps-s}) 
and pseudoscalar (\ref{wigner-foton-ps}) particle into two photons lead to two 
trigonometric inequalities (\ref{wigner-nonclassic-foton}) and (\ref{wigner-classic-foton}), 
which can be used for the experimental test of the Bohr's complementarity principle as well as 
inequality (\ref{wigner-classic-nqm}).

If we do not require $P$-parity conservation, then the decay amplitude of a pseudoscalar or scalar meson 
$P^0$ with mass $M_P$ into a photon pair has the following form:
\begin{eqnarray}
A^{P^0 \to \gamma\gamma}\, =\,\epsilon^{*\,\mu} (\lambda^{(1)})\,
                              \epsilon^{*\,\nu} (\lambda^{(2)})\,  
\left [
A\,\varepsilon_{\mu\nu\alpha\beta} k^{\alpha}_1 k^{\beta}_2\, -\, 
i\, B\, 
   \left( k_{2 \mu} k_{1 \nu}\, -\,  
          g_{\mu\nu}\,\frac{M^2_P}{2}
   \right ) 
\right ], \nonumber
\end{eqnarray}
where $A$ and $B$ are two constants, which can be calculated in the framework of QFT and may
be found for example in \cite{bb}.
For this amplitude the inequality (\ref{wigner-foton-ps}) is transformed into a trigonometric 
inequality: 
\begin{eqnarray}
\label{wigner-AB-ps}
\left (|A|^2\, -\, |B|^2 \right )\,\sin^2\theta_{ab}\,
\le\, |B|^2\, +\,\left (|A|^2\, -\, |B|^2 \right )\,  
\left(\sin^2\theta_{ac}\, +\,\sin^2\theta_{bc}\right ). 
\end{eqnarray}
The inequality (\ref{wigner-AB-ps}) can be violated only when   
$
|A| \ge \sqrt{2} |B|. 
$
The violation reaches a maximum when $|B| = 0$. In this case, inequality (\ref{wigner-AB-ps}) 
is transformed to (\ref{wigner-classic-foton}) obtained above.

The inequality (\ref{wigner-AB-ps}) is not affected if one of the final particles is a vector 
meson, instead of a photon. For example, the inequality may be written for the decay 
$B^0_d \to K^{*\, 0} \gamma$. However, in this case $|A|=|B|$. 
Therefore, the formula (\ref{wigner-AB-ps}) becomes a trivial statement: 
$|B| \ge 0$. The same trivialization of Bell's inequalities occurs in rare radiative 
decays $B^0_s \to \gamma \gamma$. In this case the cause of the triviality stems from 
these decays going through loop diagrams, which are reduced to the effective tensor operator 
$
\bar s\,\sigma^{\mu\nu} (1 + \gamma^5) b.
$ 
If the contributions from tensor and pseudotensor quark currents are essentially different,
then the inequality (\ref{wigner-AB-ps}) may be violated. This is possible, for example, in 
LR models.


\newpage
\section{Conclusion}

The following conclusions are drawn for this article.
\begin{enumerate}
   \item It is shown that the relativistic generalization of Bell's inequalities 
	 in Wigner form for the decay of a resting pseudoscalar particle into a fermion-antifermion
	 pair reproduces the nonrelativistic result (\ref{wigner-classic-nqm}) for
	 the decay of a singlet state into the two states with spin 1/2.
   \item We proved that the corrections due to small deviations from the exact antiparallel state
	 of a fermion and antifermion are quadratic by the small parameter 
         $|\vec k_1 + \vec k_2 |/M$.
   \item For the case of a scalar particle decay into a fermion-antifermion pair, we 
	 obtained a new types of Bell's inequality in Wigner form (\ref{wigner-s})
	 for the full spin correlations using the special experimental configuration (\ref{pi-popolam}).
	 It is shown that the inequality (\ref{wigner-s}) may lead to the trigonometric inequality 
	 (\ref{wigner-classic-nqm}), the same as for a pseudoscalar particle.  
   \item The decay of a scalar and pseudoscalar particle into two photons leads to new trigonometric 
     	 inequalities (\ref{wigner-nonclassic-foton}) and (\ref{wigner-classic-foton}), respectively, assuming
	 that $P$ parity is conserved. These inequalities may be experimentally tested at current colliders. 
   \item If $P$ parity is not conserved, the inequality (\ref{wigner-classic-foton}) may be 
         generalized to inequality (\ref{wigner-AB-ps}). The inequality (\ref{wigner-AB-ps}) is not affected 
         if one of the final particles is a vector meson, instead of a photon.
\end{enumerate}


\acknowledgments
The authors thank S.~Baranov, A.~Grinbaum, and J.~Metcalfe for the important 
and fruitful discussions which helped to improve the article. 
For N.~Nikitin this work was supported by a grant from the President
of the Russian Federation for Scientific Schools (1456.2008.2) and a
state contract supporting "The studies of the fundamental interactions
of the elementary particles and the computing simulations for
contemporary experiments."


\appendix


\section{A derivation of Bell's inequalities in Wigner form}
\label{app:wigner}

In this appendix we give a detailed derivation of Bell's inequalities in Wigner form 
for decays of pseudoscalar and scalar particles.


\subsection{Derivation of Bell's inequalities for a two-body decay of a pseudoscalar particle}
\label{appsub:wigner-anticorr}

Consider a case of a full anticorrelation of spin projections. It may appear, for example,
for the decay of a pseudoscalar particle into a fermion-antifermion pair (see Sec. \ref{sec:wigner2}). \\
Let us make a key assumption for the derivation. The spin projections of fermion and
antifermion on three nonparallel directions $\vec a$, $\vec b$, and $\vec c$ \textit{are 
simultaneously elements of a physical reality}. Then we can speak of non-negative
number of fermion-antifermion pairs, with  $s^{(2)}_a = +1/2$, $s^{(1)}_b = + 1/2$, 
and $s^{(1)}_c = +1/2$. Denote the number of such pairs as 
 $N(s^{(2)}_a = +1/2,\, s^{(1)}_b = + 1/2,\, s^{(1)}_c = +1/2)$.
Now it is easy to obtain the number of pairs with spin projections only on two directions:
\begin{eqnarray}
\label{N1ac}
N\left (s^{(2)}_a = +\,\frac{1}{2},\, s^{(1)}_b = +\,\frac{1}{2} \right ) &=& 
N\left (s^{(2)}_a = +\,\frac{1}{2},\, s^{(1)}_b = +\,\frac{1}{2},\, 
  s^{(1)}_c = +\,\frac{1}{2}\right )\, + \nonumber \\ 
&+& N\left (s^{(2)}_a = +\,\frac{1}{2},\, s^{(1)}_b = + \,\frac{1}{2},\, 
            s^{(1)}_c = -\,\frac{1}{2}\right ). 
\end{eqnarray}
In analogy,
\begin{eqnarray}
\label{N2ac}
N\left (s^{(2)}_a = +\,\frac{1}{2},\, s^{(1)}_c = + \,\frac{1}{2}\right ) &=& 
N\left (s^{(2)}_a = +\,\frac{1}{2},\, s^{(1)}_b = + \,\frac{1}{2},\, 
        s^{(1)}_c = +\,\frac{1}{2}\right )\, + \nonumber \\ 
&+& N\left (s^{(2)}_a = +\,\frac{1}{2},\, s^{(1)}_b = - \,\frac{1}{2},\, 
            s^{(1)}_c = +\,\frac{1}{2}\right ). 
\end{eqnarray}
And finally
\begin{eqnarray}
N\left (s^{(2)}_c = +\,\frac{1}{2},\, s^{(1)}_b = + \,\frac{1}{2}\right ) &=& 
N\left (s^{(2)}_a = +\,\frac{1}{2},\, s^{(2)}_c = + \,\frac{1}{2},\, 
        s^{(1)}_b = +\,\frac{1}{2}\right )\, + \nonumber \\ 
&+& N\left (s^{(2)}_a = -\,\frac{1}{2},\, s^{(2)}_c = + \,\frac{1}{2},\, 
            s^{(1)}_b = +\,\frac{1}{2}\right ),  \nonumber
\end{eqnarray}
or, using the anticorrelation condition (\ref{s-anticorr}) for the direction $\vec c$, 
\begin{eqnarray}
\label{N3ac}
N\left (s^{(2)}_c = +\,\frac{1}{2},\, s^{(1)}_b = + \,\frac{1}{2}\right ) &=& 
N\left (s^{(2)}_a = +\,\frac{1}{2},\, s^{(1)}_b = + \,\frac{1}{2},\, 
        s^{(1)}_c = -\,\frac{1}{2}\right )\, + \nonumber \\ 
&+& N\left (s^{(2)}_a = -\,\frac{1}{2},\, s^{(1)}_b = + \,\frac{1}{2},\, 
            s^{(1)}_c = -\,\frac{1}{2}\right ). 
\end{eqnarray}
In equalities (\ref{N1ac}) -- (\ref{N3ac}) each term on the right-hand
side is non-negative. 
Hence, we can write a kind of ``triangle inequality'':
\begin{eqnarray}
\label{Triang-ac}
N\left (s^{(2)}_a = +\,\frac{1}{2},\, s^{(1)}_b = + \,\frac{1}{2} \right )\,& \le & 
N\left (s^{(2)}_a = +\,\frac{1}{2},\, s^{(1)}_c = + \,\frac{1}{2}\right )\, +\nonumber\\ 
&+& N\left (s^{(2)}_c = +\,\frac{1}{2},\, s^{(1)}_b = + \,\frac{1}{2}\right ). 
\end{eqnarray}
Since the number of fermion-antifermion pairs is inversely proportional to the decay
probability, we immediately have the inequality (\ref{wigner-ps}).

The basic inequality (\ref{Triang-ac}) may be rewritten in a few equivalent forms. For example,
if $\vec b$ and $\vec c$ are changed to opposite directions, then using the condition 
(\ref{s-anticorr}) one can obtain the inequality
\begin{eqnarray}
N\left (s^{(2)}_a = +\,\frac{1}{2},\, s^{(1)}_b = - \,\frac{1}{2} \right )\,& \le & 
N\left (s^{(2)}_a = +\,\frac{1}{2},\, s^{(1)}_c = - \,\frac{1}{2}\right )\, +\nonumber\\ 
&+& N\left (s^{(2)}_b = +\,\frac{1}{2},\, s^{(1)}_c = + \,\frac{1}{2}\right ).\nonumber 
\end{eqnarray}
If we change only the direction $\vec c$ to the opposite, then the the following
inequality appears: 
\begin{eqnarray}
N\left (s^{(2)}_a = +\,\frac{1}{2},\, s^{(1)}_b = + \,\frac{1}{2} \right )\,& \le & 
N\left (s^{(2)}_a = +\,\frac{1}{2},\, s^{(1)}_c = - \,\frac{1}{2}\right )\, +\nonumber\\ 
&+& N\left (s^{(2)}_c = -\,\frac{1}{2},\, s^{(1)}_b = + \,\frac{1}{2}\right ), \nonumber 
\end{eqnarray}
It can be weakened by rewriting it in the following form:
\begin{eqnarray}
&& N\left (s^{(2)}_a = +\,\frac{1}{2},\, s^{(1)}_b = + \,\frac{1}{2} \right ) \, + \,  
N\left (s^{(2)}_a = +\,\frac{1}{2},\, s^{(1)}_c = + \,\frac{1}{2}\right )\, +\nonumber\\ 
&&\qquad\qquad +\, N\left (s^{(2)}_c = +\,\frac{1}{2},\, s^{(1)}_b = + \,\frac{1}{2}\right )
\,\le\, N_{tot}. \nonumber
\end{eqnarray}
If the vectors $\vec a$, $\vec b$, and $\vec c$ lie in the same plane, then in the framework of NQM,
the last inequality is reduced to the following trigonometric inequality:
\begin{eqnarray}
\sin^2\frac{\theta_{ab}}{2}\, + \,\sin^2\frac{\theta_{ac}}{2}\, +
                                \,\sin^2\frac{\theta_{bc}}{2}\,\le\, 2, \nonumber
\end{eqnarray}
which is violated when the angle between the vectors $\vec a$ and $\vec b$ is close to $\pi$, and
the direction $\vec c$ bisects this angle. This condition is more
strict than the condition of violation of the inequality (\ref{wigner-classic-nqm}).

Since all the preceding variants of Bell's inequalities in Wigner form
are either equivalent to the inequality (\ref{Triang-ac}) or weaker, we consider 
only the relativistic generalization of  (\ref{Triang-ac}).


\subsection{Derivation of Bell's inequalities for the decay of a scalar particle into two fermions}
\label{appsub:wigner-corr}

Consider the case of a full spin projection correlation. The pure full
correlation does not appear
in the decays. Usually there are contributions with correlation and with anticorrelation as well.
In this case there are much fewer possibilities for the violation of Bell's inequalities in QFT than
in the cases of the pure full correlation or of the pure full anticorrelation. However, as was shown
in the Sec. \ref{sec:wigner} with use the experimental configuration \ref{pi-popolam}, QFT predicts 
the full spin correlation, for example, for the case of the decay of a scalar particle.
As in Appendix \ref{appsub:wigner-anticorr}, we assume that the spin projections of
the fermion and antifermion on three nonparallel directions set by unitary vectors $\vec a$, 
$\vec b$, and $\vec c$ are simultaneously elements of a physical reality. Then
\begin{eqnarray}
\label{N1c}
N\left (s^{(2)}_a = +\,\frac{1}{2},\, s^{(1)}_b = -\,\frac{1}{2} \right ) &=& 
N\left (s^{(2)}_a = +\,\frac{1}{2},\, s^{(1)}_b = -\,\frac{1}{2},\, 
  s^{(1)}_c = +\,\frac{1}{2}\right )\, + \nonumber \\ 
&+& N\left (s^{(2)}_a = +\,\frac{1}{2},\, s^{(1)}_b = - \,\frac{1}{2},\, 
            s^{(1)}_c = -\,\frac{1}{2}\right ). 
\end{eqnarray}
In analogy,
\begin{eqnarray}
\label{N2c}
N\left (s^{(2)}_a = +\,\frac{1}{2},\, s^{(1)}_c = - \,\frac{1}{2}\right ) &=& 
N\left (s^{(2)}_a = +\,\frac{1}{2},\, s^{(1)}_b = + \,\frac{1}{2},\, 
        s^{(1)}_c = -\,\frac{1}{2}\right )\, + \nonumber \\ 
&+& N\left (s^{(2)}_a = +\,\frac{1}{2},\, s^{(1)}_b = - \,\frac{1}{2},\, 
            s^{(1)}_c = -\,\frac{1}{2}\right ). 
\end{eqnarray}
And finally
\begin{eqnarray}
N\left (s^{(2)}_c = +\,\frac{1}{2},\, s^{(1)}_b = - \,\frac{1}{2}\right ) &=& 
N\left (s^{(2)}_a = +\,\frac{1}{2},\, s^{(2)}_c = + \,\frac{1}{2},\, 
        s^{(1)}_b = -\,\frac{1}{2}\right )\, + \nonumber \\ 
&+& N\left (s^{(2)}_a = -\,\frac{1}{2},\, s^{(2)}_c = + \,\frac{1}{2},\, 
            s^{(1)}_b = -\,\frac{1}{2}\right ), \nonumber
\end{eqnarray}
or, considering the correlations of spin projections on the direction $\vec c$,
\begin{eqnarray}
\label{N3c}
N\left (s^{(2)}_c = +\,\frac{1}{2},\, s^{(1)}_b = - \,\frac{1}{2}\right ) &=& 
N\left (s^{(2)}_a = +\,\frac{1}{2},\, s^{(1)}_b = - \,\frac{1}{2},\, 
        s^{(1)}_c = +\,\frac{1}{2}\right )\, + \nonumber \\ 
&+& N\left (s^{(2)}_a = -\,\frac{1}{2},\, s^{(1)}_b = - \,\frac{1}{2},\, 
            s^{(1)}_c = +\,\frac{1}{2}\right ). 
\end{eqnarray}
From equalities (\ref{N1c}) -- (\ref{N3c}) follows the inequality
\begin{eqnarray}
\label{Triang-c}
N\left (s^{(2)}_a = +\,\frac{1}{2},\, s^{(1)}_b = - \,\frac{1}{2} \right )\,& \le & 
N\left (s^{(2)}_a = +\,\frac{1}{2},\, s^{(1)}_c = - \,\frac{1}{2}\right )\, +\nonumber\\ 
&+& N\left (s^{(2)}_c = +\,\frac{1}{2},\, s^{(1)}_b = - \,\frac{1}{2}\right ) 
\end{eqnarray}
and its probabilistic analog, the inequality (\ref{wigner-s}). Like formula
(\ref{Triang-ac}), the inequality (\ref{Triang-c}) can be transformed to other equivalent 
inequalities by switching the directions of $\vec b$ and $\vec c$.


\section{A relativistic spin 1/2 operator and solutions for a free Dirac equality}
\label{app:spin}

Let the free Dirac particle of mass $m$ propagate in the laboratory coordinate system over the direction
defined by a unitary vector,
\begin{eqnarray}
\label{n-1v}
\vec n\, =\, 
\bigl (
\sin\theta\, \cos\phi,\, \sin\theta\,\sin\phi,\,\cos\theta
\bigr ), 
\end{eqnarray}
where $\theta\,\in\,\left [0,\,\pi \right )$, $\phi\,\in\,\left [0,\, 2\,\pi \right )$. 
In this coordinate system, the particle has energy  $\varepsilon_p$ and momentum
$\vec p\, =\, |\vec p\, |\,\vec n$.

The solution of the free Dirac equation in the standard representation for the particle has the form
\begin{eqnarray}
\label{dirac_u}
u(\vec p,\, s_a,\, \vec a\, )\, =\,
\left (
\begin{array}{lc}
\sqrt{\varepsilon_p + m}                                    & \chi_{s_a}(\vec a\, ) \\
\sqrt{\varepsilon_p - m}\,\left(\vec\sigma\vec n\,\right)   & \chi_{s_a}(\vec a\, )     
\end{array}
\right ) 
\end{eqnarray}
and for the antiparticle has the form
\begin{eqnarray}
\label{dirac_v}
v(\vec p,\, s_a,\, \vec a\, )\, =\,
\left (
\begin{array}{lc}
\sqrt{\varepsilon_p - m}\,\left(\vec\sigma\vec n\,\right)   & \xi_{-s_a}(\vec a\, ) \\
\sqrt{\varepsilon_p + m}                                    & \xi_{-s_a}(\vec a\, )   
\end{array}
\right ), 
\end{eqnarray}
where $s_a = \pm 1/2$ is a spin projection on a unitary vector direction 
\begin{eqnarray}
\vec a\, =\, 
\left (
\sin\theta_a\, \cos\phi_a,\, \sin\theta_a\,\sin\phi_a,\,\cos\theta_a
\right ).\nonumber
\end{eqnarray}
Two-component spinors $\chi_{s_a}(\vec a\, )$ and $\xi_{-s_a}(\vec a\, )$ obey the
normalization conditions 
$
\chi_{s_a}(\vec a\, )^{\dagger} \chi_{s'_a}(\vec a\, ) = 
\delta_{s_a\, s'_a}
$
and 
$
\xi_{-s_a}(\vec a\, ) = -\, 2\, s_a\, \chi_{-s_a}(\vec a\, ).
$

The solution (\ref{dirac_u}) must be an eigenfunction of a projection operator
\begin{eqnarray}
\left (\vec a\, \vec O\, \right )\, u(\vec p,\, s_a,\, \vec a\, )\, =\,
2\, s_a\, u(\vec p,\, s_a,\, \vec a\, ),  
\end{eqnarray}
corresponding to eigenvalues $2 s_a = \pm 1$. The operator $\vec O$ is a relativistic generalization
of a spin 1/2 operator for a free particle and can be written as \cite{stech}
\begin{eqnarray}
\label{spinO}
\vec O\, =\, -\,\gamma^5\,\vec\gamma\, +\,\gamma^5\,\frac{\vec p}{\varepsilon_p}\, +\, 
\frac{\vec p\,\gamma_5\, (\vec\gamma,\,\vec p\, )}{\varepsilon_p\, (\varepsilon_p + m)}, 
\end{eqnarray}
where in the standard representation the matrix $\gamma^5$ can be
written as
$$
\gamma^5\, =\, i \gamma^0 \gamma^1 \gamma^2 \gamma^3\, =\, 
\left (
   \begin{array}{cccc}
    0 & 0 & 1 & 0 \\
    0 & 0 & 0 & 1 \\
    1 & 0 & 0 & 0 \\
    0 & 1 & 0 & 0
   \end{array}
   \right ).
$$
Components of the operator (\ref{spinO}) satisfy the standard commutation relations 
for doubled components of nonrelativistic spin-1/2 operator ($\epsilon^{123} = + 1$),
\begin{eqnarray}
\left [ O^i,\, O^j \right ]\, =\, 2\, i\,\epsilon^{ijk}\, O^k .\nonumber
\end{eqnarray}
This allows one to test the Bohr's complimentarity principle in QFT in analogy with NQM.

From the explicit form of the operator $\vec O$ follows formulas for two-component spinors:
\begin{eqnarray}
&& \chi_{s_a\, =\, +\, 1/2}(\vec a\, )\, \equiv\,\chi_+(\vec a\, )\, =\,
   \left (
   \begin{array}{l}
     \cos\,\frac{\theta_a}{2}\, e^{- i \phi_a/2} \\
     \sin\,\frac{\theta_a}{2}\, e^{i \phi_a/2}
   \end{array}
   \right ), \nonumber \\
&& \chi_{s_a\, =\, -\, 1/2}(\vec a\, )\, \equiv\,\chi_-(\vec a\, )\, =\,
   \left (
   \begin{array}{l}
     -\,\sin\,\frac{\theta_a}{2}\, e^{- i \phi_a/2} \\
     \quad\,\cos\,\frac{\theta_a}{2}\, e^{i \phi_a/2}
   \end{array}
   \right ). \nonumber
\end{eqnarray}
Hence, when $\phi_a = \phi_b = 0$ 
\begin{eqnarray}
\label{formuli-chi}
&&\chi^{\dagger}_+(\vec a\, )\,\chi_-(\vec b\, )\, =\, \sin\,\frac{\theta_{ab}}{2},\nonumber\\
&&\vec w_{++}\, =\,\chi^{\dagger}_+(\vec a\, )\,\vec\sigma\,\chi_-(\vec b\, )\, =\,
  \left (
          \cos\,\frac{\kappa_{ab}}{2},\, 
          -\, i\, \cos\,\frac{\theta_{ab}}{2},\, 
          -\, \sin\,\frac{\kappa_{ab}}{2}
  \right ),\\
&&\vec w_{+-}\, =\,\chi^{\dagger}_+(\vec a\, )\,\vec\sigma\,\chi_+(\vec b\, )\, =\,
  \left (
          \sin\,\frac{\kappa_{ab}}{2},\, 
       i\,\sin\,\frac{\theta_{ab}}{2},\, 
          \cos\,\frac{\kappa_{ab}}{2}
  \right ), 
\nonumber
\end{eqnarray}
where $\theta_{\alpha \beta} = \theta_{\alpha} - \theta_{\beta}$, 
    $\kappa_{\alpha \beta} = \theta_{\alpha} + \theta_{\beta}$, 
and $\{\alpha, \beta \} = \{a, b, c\}$.



\end{document}